\documentclass[letterpaper, 10 pt, conference]{ieeeconf}  

\IEEEoverridecommandlockouts                              

\usepackage{soul}

\newcommand\topic[1]{\textcolor{orange}{}}  

\newcommand{\dtau}{\ensuremath{\dot{\tau}}}
\newcommand{\alphamin}{\ensuremath{\alpha^{-}}}
\newcommand{\alphamind}{\ensuremath{\alpha_\textrm{d}^{-}}}
\newcommand{\fdm}{\ensuremath{f_{\textrm{dm}}}}
\newcommand{\te}{\ensuremath{t_\textrm{e}}}
\newcommand{\tei}{\ensuremath{t_{\textrm{e},1}}}
\newcommand{\teii}{\ensuremath{t_{\textrm{e},2}}}
\newcommand{\temax}{\ensuremath{t_{\textrm{e,max}}}}
\newcommand{\Tdm}{\ensuremath{\Delta T_{\textrm{dm}}}}
\newcommand{\Tmax}{\ensuremath{T_{\textrm{max}}}}
\newcommand{\Twt}{\ensuremath{T_{\textrm{wait}}}}
\newcommand{\vped}{\ensuremath{v_{\textrm{ped}}}}
\newcommand{\sped}{\ensuremath{s_{\textrm{ped}}}}

\newcommand{\sei}{\ensuremath{s_{\textrm{e},1}}}
\newcommand{\seii}{\ensuremath{s_{\textrm{e},2}}}

\newcommand{\velei}{\ensuremath{v_{\textrm{e},1}}}

\newcommand{\accei}{\ensuremath{a_{\textrm{e},1}}}

\newcommand{\jei}{\ensuremath{j_{\textrm{e},1}}}

\newcommand{\tauinit}{\ensuremath{\tau_\textrm{init}}}
\newcommand{\wte}{\ensuremath{w_{\te}}}
\newcommand{\wTBv}{\ensuremath{w_{\textrm{TB,v}}}}
\newcommand{\wTBp}{\ensuremath{w_{\textrm{TB,p}}}}
\newcommand{\wWTp}{\ensuremath{w_{\textrm{WT}}}}
\newcommand{\wj}{\ensuremath{w_{j}}}
\newcommand{\ccomfv}{\ensuremath{C_{\textrm{comf,v}}}}
\newcommand{\cutilv}{\ensuremath{C_{\textrm{util,v}}}}
\newcommand{\cutilp}{\ensuremath{C_{\textrm{util,p}}}}
\newcommand{\cjoint}{\ensuremath{C_{\textrm{joint}}}}

\newcommand{\wu}{\ensuremath{w_{u}}}
\newcommand{\xiopt}{\ensuremath{\xi^{*}}}
\newcommand{\ve}[1]{\ensuremath{\boldsymbol{\mathrm{#1}}}}

\newcommand{\figref}[1]{Fig.~\ref{#1}}
\newcommand{\tabref}[1]{Tab.~\ref{#1}}
\newcommand{\secref}[1]{Sec.~\ref{#1}}
\usepackage{cite}
\usepackage[T1]{fontenc}
\usepackage{amsmath,amssymb,amsfonts}
\usepackage{algorithmic}
\usepackage{bigints}
\usepackage{graphicx}
\usepackage{afterpage}
\usepackage{placeins}
\usepackage[caption=false]{subfig}
\usepackage{dblfloatfix}
\usepackage{epsfig}
\usepackage{textcomp}
\usepackage{makecell}
\usepackage{xcolor, soul}
\usepackage[nolist,nohyperlinks]{acronym}
\usepackage[acronym]{glossaries}
\usepackage{pdfpages}
\usepackage{svg}

\usepackage{tikz}
\usepackage[color=black,opacity=1,angle=0,scale=1]{background}
\backgroundsetup{
	contents={
		\begin{tikzpicture}
			\node at (current page.center) [align=center] {\textcopyright 2025 IEEE. Personal use of this material is permitted. \\ Permission from IEEE must be obtained for all other uses. This work is licensed under a Creative Commons Attribution 4.0 International License (CC BY 4.0)
			\\ DOI: 10.1109/IV64158.2025.11097656 };
	\end{tikzpicture}},
	placement=bottom,
	scale=0.6,
	vshift=20
}

\makeglossaries

\makeatletter
\g@addto@macro\bfseries{\boldmath}
\makeatother

\title{Optimal Behavior Planning for Implicit Communication using a Probabilistic Vehicle-Pedestrian Interaction Model
	\thanks{This research was supported by the HEIDI project which has received funding from the European Union's Horizon Europe research and innovation programme under Grant Agreement No. 101069538. Views and opinions expressed are those of the authors only and do not necessarily reflect those of the European Union or CINEA. Neither the European Union nor the granting authority can be held responsible for them.}
}

\author{Markus Amann$^{1, 2}$, Malte Probst$^{1}$, Raphael Wenzel$^{1}$, Thomas H. Weisswange$^{1}$, Miguel \'{A}ngel Sotelo$^{2}$
	\thanks{$^{1}$Honda Research Institute Europe GmbH, Carl-Legien-Str. 30, 63073 Offenbach, Germany, Email:{\tt\footnotesize \{firstname.lastname\}@honda-ri.de}}%
	\thanks{$^{2}$Computer Engineering Department, University of Alcal\'{a}, 28805 Alcal\'{a} de Hernares, Madrid, Spain, Email:\, {\tt\footnotesize miguel.sotelo@uah.es}}%
}

\begin{document}
	\bstctlcite{IEEEexample:BSTcontrol}
	\maketitle
	\thispagestyle{empty}
	\pagestyle{empty}

	\begin{abstract}
		In interactions between \acrfull{avs} and crossing pedestrians, modeling implicit vehicle communication is crucial. In this work, we present a combined prediction and planning approach that allows to consider the influence of the planned vehicle behavior on a pedestrian and predict a pedestrian's reaction. We plan the behavior by solving two consecutive \acrfull{ocps} analytically, using variational calculus. We perform a validation step that assesses whether the planned vehicle behavior is adequate to trigger a certain pedestrian reaction, which accounts for the closed-loop characteristics of prediction and planning influencing each other. In this step, we model the influence of the planned vehicle behavior on the pedestrian using a probabilistic behavior acceptance model that returns an estimate for the crossing probability. The probabilistic modeling of the pedestrian reaction facilitates considering the pedestrian's costs, thereby improving cooperative behavior planning. We demonstrate the performance of the proposed approach in simulated vehicle-pedestrian interactions with varying initial settings and highlight the decision making capabilities of the planning approach.
	\end{abstract}

	\section{Introduction}
	Many traffic situations require communication between interaction partners to coordinate their actions. Communication can happen in various ways and through several modalities \cite{habibovic.2018, dietrich2.2020, eisele.2024}. Implicit communication by means of the vehicle behavior is especially relevant in interactions between pedestrians and \acrfull{avs}, since driver-focused communication cues such as eye contact are no longer available \cite{dietrich2.2020}. Moreover, recent studies suggest that communication through explicit \acrfull{ehmis} needs to be aligned with the implicit communication exhibited through the vehicle behavior \cite{martin.2024, eisele.2024}. Communication is bidirectional which means that a traffic participant does not only communicate its own intention, but it also needs to interpret the behavior of the interaction partner to predict their intention. In addition to predicting the intention, it is crucial to model the influence of the own behavior on the interaction partner and estimate their reaction to the planned behavior \cite{mertens.2020}. Since the planned behavior influences the evolution of the scenario, it is essential to understand the implications of the current behavior for future planning and predictions. The closed-loop characteristics of prediction and behavior planning influencing each other is a well known paradigm in interaction-aware and cooperative behavior planning \cite{hagedorn.2024, wenzel.2021, burger.2022}. 
	
\begin{figure}[t]
		\centering
		\includegraphics[width=1.\linewidth]{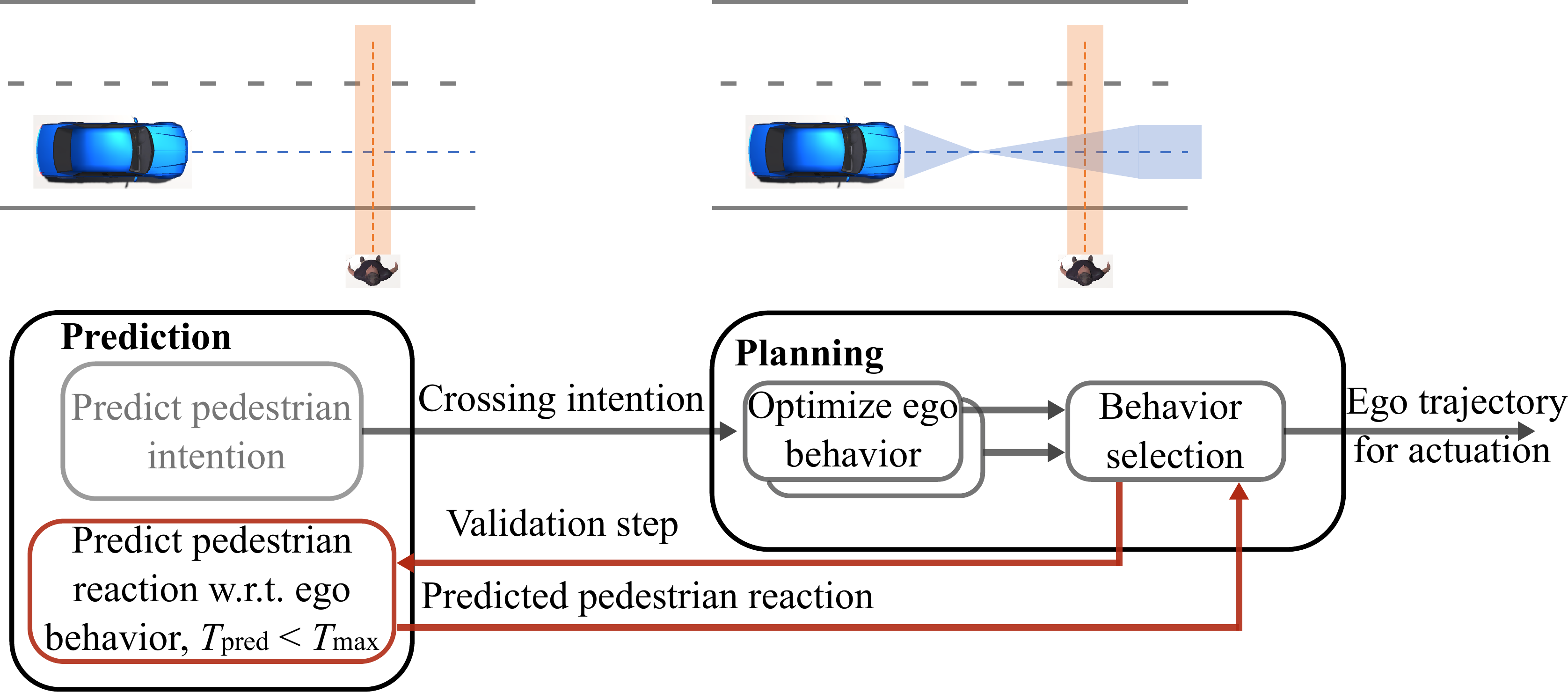}
		\caption{Architecture of combined prediction and planning approach. Classical process in black. Additional modules comprising behavior validation step for implicit communication modeling in red. Pedestrian intention estimation (light gray box) is assumed to be given in this work.}
		\label{fig:bilevel_prediction}
		\vspace*{-3mm}
	\end{figure}

	In situations where a pedestrian intending to cross the road interacts with a vehicle, the vehicle behavior has a significant impact on the crossing decision \cite{schneemann.2016}. In order to induce crossing behavior, the vehicle's yielding intention needs to be apparent to the pedestrian. In this work, we research vehicle-pedestrian interactions and model the influence of implicit communication by means of the vehicle behavior on a pedestrian's crossing decision. We propose a novel behavior planning approach that allows considering the effect of the planned behavior on the pedestrian and predict the pedestrian's reaction. We perform a dense parameter sampling over possible end positions and end times for the vehicle to reach during the interaction and solve an \acrfull{ocp} to plan the vehicle behavior. We couple prediction and planning by estimating the pedestrian reaction to the planned behavior for each sampled trajectory. Based on the estimated reaction, we adjust a joint cost function. We refer to the pedestrian reaction modeling and the coupling of behavior planning and prediction as validation step that assesses whether a planned vehicle behavior is adequate to trigger a certain pedestrian behavior. This allows selecting the most suitable behavior for an interaction. \figref{fig:bilevel_prediction} depicts an exemplary vehicle-pedestrian interaction and the architecture of the proposed approach.
	
	\section{Related Work}
	
	Pedestrian behavior modeling is an important and ongoing field of research. Recent studies about vehicle-pedestrian interactions demonstrate the influence and highlight the importance of the vehicle behavior on the crossing decision of pedestrians \cite{martin.2024, dietrich2.2020, dietrich.2020, schneemann.2016, schroeder.2014, pekkanen.2022, tian.2023}. As shown in \cite{pekkanen.2022, tian.2023, schroeder.2014}, pedestrians follow a two-stage decision making process and exhibit behavior that results in a bimodal distribution of crossing events. In early phases of the interaction, when the vehicle is at a large distance, pedestrians make their crossing decision mainly based on the time gap. If this time is not large enough and the pedestrian rejects to cross in the first stage, a pedestrian will only cross if the vehicle shows pronounced yielding signs. This means that in later phases, when the vehicle gets closer, pedestrians focus their crossing decision on the changes in vehicle behavior. Evidence accumulation models have shown to be a promising approach to capture the integrating nature of the crossing decision process \cite{theisen.2024, kalantari.2023, pekkanen.2022}. These models assume that pedestrians repeatedly make a crossing decision and accumulate evidence for a yielding intention of the vehicle over time. 
	
	Pedestrian predictions play a crucial role for planning suitable behavior in traffic interactions \cite{camara.2021}. In \cite{bonnin.2014}, the authors propose a framework that incorporates contextual information as an extension to a generic pedestrian crossing prediction model. They argue that conventional generic models tend to neglect important context specific features. Incorporating such information allows for more accurate predictions in certain situations. However, focusing on these features carries the risk of overpredicting crossing events. In \cite{melo.2023}, the authors present a neurosymbolic framework that predicts pedestrian intention together with an explanation based on visual input. This allows the vehicle to adapt its behavior to the predicted intention and react accordingly. However, the prediction horizon for which such neural network based models work reliably is often too short to proactively plan vehicle behaviors \cite{schoeller.2020}. 
	
	In addition to providing short-term intention predictions, it is necessary to approximate the pedestrian behavior over the whole interaction horizon. In \cite{wenzel.2021}, the authors model interactions between two vehicles relative to a shared traffic space and plan the ego behavior for the subjectively best passing order with respect to a joint cost function assuming that the interaction partner will concur. This approach allows evaluating multiple different resolutions and selecting the best outcome. However, assuming that the interaction partner will plan their behavior according to the best resolution neglects the consequences of implicit communication and the details of the influence of the ego behavior on the interaction partner. In \cite{burger.2022}, the authors model the interaction between two vehicles as a Stackelberg Game and solve the game by means of a nested optimization problem. The alternating turn-taking allows planning the own behavior to achieve certain reactions. In \cite{siebinga.2023}, the authors present a risk-based planning framework that maintains a belief over the future position of the other agent in interactions between two vehicles. The belief gets constantly updated by the actual position of the interaction partner, which allows modeling the received communication. Yet, the approach does not consider a mapping of the own communication modalities to the own belief over the assumed reaction to the communication. In \cite{skugor.2023}, the authors use a gap acceptance model to consider the crossing probability of a pedestrian based on the time gap between the ego vehicle and the pedestrian. However, this approach omits the importance of the vehicle yielding behavior on the crossing decision.  
	
	Being aware of the influence of the own behavior on the behavior of the interaction partner is an important step in modeling vehicle-pedestrian interactions, especially when it comes to the communication between the interaction partners. The own behavior inevitably entails some information about the own intention, even if no communication is intended \cite{mertens.2020}. In \cite{eisele.2024}, the authors argue that not activating an \acrshort{ehmi} that informs a pedestrian about the vehicle state or intention can communicate the vehicle's non-yielding intention. In previous work, we demonstrated the possible benefit of communicating the vehicle's intention of giving right of way using a comprehensible \acrshort{ehmi} \cite{amann.2024}. Whether implicitly through the vehicle's behavior or explicitly by means of an \acrshort{ehmi}, it is essential to carefully plan the vehicle's actions with respect to the communication.

	\section{Methodology}
	\subsection{Vehicle Model}
	In this work, we assume that the vehicle moves along a straight path which allows us to neglect the lateral vehicle dynamics. We use a 4th-order point mass model to describe the longitudinal vehicle dynamics. The vehicle state $\ve{x}(t)\in \mathbb{R}^n$ with $n=4$ entails the vehicle's position $s(t)$, longitudinal velocity $v(t)$, longitudinal acceleration $a(t)$, and longitudinal jerk $j(t)$. The control input is given by the first derivative of jerk $\dot{j}(t)$. The vehicle dynamics are described as
	\begin{equation}
		\dot{\ve{x}}(t) = \begin{bmatrix}
			v(t), & a(t), & j(t), & u(t)
		\end{bmatrix}^T \, \textrm{with} \, u(t) = \dot{j}(t)\,.
		\label{eq:state_model}
	\end{equation}
	For the sake of readability, we omit the time dependent notation of the dynamic variables, unless explicitly needed.
	\subsection{\acrfull{OCP} Formulation}
	We target the planning task by formulating a comfort and efficiency optimized \acrshort{ocp} and solve the problem using variational calculus. Solving \acrshort{ocps} by means of variational calculus is a commonly used approach that allows to derive certain optimality conditions and boundary conditions for the states at the end of the optimization interval, thereby transforming the \acrshort{ocp} into a boundary value problem \cite{papageorgiou.2015}. Such problems can often be solved analytically, which is why approaches based on variational calculus are often refered to as \textit{first-optimize-then-discretize} approaches \cite{papageorgiou.2015}. Equations \eqref{eq:objective_fun}--\eqref{eq:bc_end_conditions} describe the general \acrshort{ocp} of the planning task. 
	\begin{align}
		\underset{u}{\min} & \quad \wte\te + \int_{t_0}^{\te} \frac{\wj}{2}j^2 + \frac{\wu}{2}\dot{j}^2 \textrm{d}t \label{eq:objective_fun}\\
		\textrm{s.t.} & \quad \dot{\ve{x}} = \ve{f}(\ve{x}, u), \qquad \ve{x}(t_0) = \ve{x}_0 \label{eq:bc_system_dynamics}\\
		& \quad \ve{g}(\ve{x}(\te), \te) = 0 \label{eq:bc_end_conditions}
	\end{align}
	Equation \eqref{eq:objective_fun} denotes the cost function that describes a trade-off between driving comfort (penalization of $j$), energy consumption (penalization of $\dot{j}$), and time efficiency (penalization of $\te$). Equation \eqref{eq:bc_system_dynamics} assures that the trajectories fulfill the system dynamics and the initial conditions. Equation \eqref{eq:bc_end_conditions} guarantees the fulfillment of the terminal equality conditions $\ve{g}(\ve{x}(\te), \te)$ for those end states fixed to desired values at the end of the optimization. This also applies to the end time $\te$, which can either be fixed or left free as an optimization variable\footnote[1]{It is reasonable to \textit{either} set $\te$ to a desired value by means of a dedicated terminal condition \textit{or} leave $\te$ free penalizing it with a non-zero weight. Doing both at the same time adds a constant offset to the cost function.}. 
	Solving the \acrshort{ocp} leads to the inhomogeneous linear second-order differential equation 
	\begin{equation}
		\ddot{j} = \frac{\wj}{\wu}j + \frac{c_1}{2\wu}t^2 - \frac{c_2}{\wu}t + \frac{c_3}{\wu} \, .
	\end{equation}
	This equation can be solved analytically resulting in the solution of the optimal jerk trajectory 
	\begin{equation}
		j^* = k_1e^{\sqrt{\frac{\wj}{\wu}}t} + k_2e^{-\sqrt{\frac{\wj}{\wu}}t} - \frac{c_1 t^2}{2\wj} + \frac{c_2 t}{\wj} - \frac{c_3}{\wj} - \frac{c_1\wu}{\wj^2} \label{eq:optimal_jerk}
	\end{equation}
	with $k_1$, $k_2$, $c_1$, $c_2$, and $c_3$ being integration constants that can be determined by solving the system of non-linear equations resulting from the boundary conditions. All other optimal state and control input trajectories can be calculated by derivation and integration based on \eqref{eq:optimal_jerk}.
	\subsection{Separation of the \acrshort{ocp} using Space-Time Sampling}
	We separate the overall planning task of the whole interaction into two subproblems with different boundary conditions with each problem following the general \acrshort{ocp} \eqref{eq:objective_fun}--\eqref{eq:bc_end_conditions}. The first subproblem focuses on resolving the interaction between the vehicle and the pedestrian by means of implicit communication. It is used to find a solution that is likely to trigger a desired pedestrian reaction. The second subproblem, which starts at the end of the first subproblem, is used to plan the vehicle behavior for the remaining distance up to the pedestrian's crossing position. The crossing position $\sped$ denotes the constant position along the vehicle path at which the pedestrian intends to cross the road. For each subproblem, the optimal trajectory $\xiopt$ can be calculated according to \eqref{eq:optimal_jerk} and the integration constants can be determined based on the respective boundary conditions. 
	
	We apply a space-time grid sampling strategy to solve the first subproblem and find suitable vehicle behavior that resolves the interaction. We sample the vehicle's end position $\sei \in (s_0, \sped]$ and fixed end time $\tei \in (t_0, \temax]$ creating a large grid of possible positions and future times at which the vehicle could be at the end of the first subproblem. The computational effort of solving the resulting \acrshort{OCP} for each grid point depends on the sampling step sizes $\Delta\sei$ and $\Delta\tei$ as well as the lengths of the intervals $\sped$ and $\temax$. A thought-out sampling strategy, e.g. reducing the sampling density at the edges of the grid, can help reducing the computational intensity. Since $\tei$ is fixed, $\wte$ is set to zero and the terminal conditions \eqref{eq:bc_end_conditions} guarantee the fulfillment of $\tei$ at the end of the first subproblem. In addition to $\sei$ and $\tei$, vehicle states $\accei$ and $\jei$ are fixed and set to zero. The vehicle velocity $\velei$ at the end of the first subproblem is free and results implicitly from solving the \acrshort{ocp}. In other words, the \acrshort{ocp} of the first subproblem brings the vehicle to a certain position within a certain desired time with free velocity without any residual acceleration and jerk. Due to the independent sampling of $\tei$ and $\sei$, the \acrshort{ocp} of the first subproblem is solved for combinations that might lead to trajectories that are infeasible to actuate for the vehicle. Hence, after solving the \acrshort{ocp} for a certain $\tei$-$\sei$-combination, the solution is passed to a feasibility filter to make sure that the trajectory does not exceed any state or input limitations. 
	
	After solving the \acrshort{ocp} of the first subproblem for all $\tei$-$\sei$-combinations, the \acrshort{ocp} of the second subproblem is solved similarly for each combination. The general solution is the same as for the first subproblem, except for the boundary conditions. To assure smooth trajectories at the transition between the two subproblems the terminal state at the end of the first subproblem is used as the initial state for the second subproblem. For the second subproblem, all terminal vehicle states are left free except for the end position $\seii$ which is set to $\sped$ such that the optimization ends at the position where the vehicle passes the pedestrian. The end time $\teii$ is left free (and penalized), which, together with the free end states, allows the optimization to find the optimal time and vehicle behavior to continue driving. 
	
	\subsection{Pedestrian Behavior Estimation}\label{subsec:pedestrian_model}
	According to \cite{pekkanen.2022}, the change rate of the time gap, $\dtau$, represents the vehicle behavior by means of distance, vehicle speed, and vehicle acceleration. The change rate $\dtau$ can be determined as the derivative of the time gap $\tau$ over time 
	\begin{align}
		\dtau(t) = \frac{\partial \tau(t)}{\partial t} &= \frac{\partial \big[\big(\sped - s(t)\big) / v(t)\big]}{\partial t} \nonumber \\
		&= \frac{-a(t)\big(\sped - s(t)\big)}{v^2(t)}-1 \, .
	\end{align}
	The change rate $\dtau$ can be interpreted as a proxy variable characterizing the future evolution of the vehicle behavior. Observing $\dtau$ allows approximating the remaining time until the vehicle reaches the pedestrian. Models that estimate the pedestrian behavior only based on the time gap $\tau$ are suitable for modeling the decision making in interactions with vehicles driving at constant speed \cite{tian.2023, petzoldt.2014}. Yet, these models can not explain all types of behavior. If a pedestrian decides to reject an initial time gap $\tauinit$ according to their gap acceptance model, $\tau$ needs to increase over the course of the interaction in order to create a suitable time gap which is sufficient for the pedestrian. However, even yielding behavior in which the vehicle applies a constant deceleration to come to a full stop right in front of the pedestrian (target braking) will not result in a larger gap than $\tauinit$. Target braking represents the minimal safe behavior to avoid a collision with the pedestrian. This behavior corresponds to a constant $\dtau = -\frac{1}{2}$ which continuously decreases the time gap until the vehicle reaches the pedestrian and eventually comes to a full stop opening an infinitely large time gap. In \cite{tian.2023}, the authors demonstrate the impact of yielding behavior on the pedestrian's crossing decision, which was also found for target braking maneuvers. 
	Based on the finding of \cite{pekkanen.2022, tian.2023}, we propose a probabilistic behavior acceptance model that accounts for the influence of changes in vehicle behavior on the pedestrian's crossing decision, facilitating cooperation through anticipatory behavior planning. We use two probabilistic models $\Phi(\tau)$ and $\Psi(\dtau)$ to estimate the probability of accepting a certain time gap $\tau$ or vehicle behavior $\dtau$, respectively. As denoted in \eqref{eq:phi_tau} and \eqref{eq:psi_dtau}, each model is described by a sigmoid function of the respective parameter
	\begin{align}
		\Phi(\tau) &= \frac{1}{1+\exp\big(-1.2(\tau-5)\big)} \label{eq:phi_tau}\\
		\Psi(\dtau) &= \frac{1}{1+\exp\big(-1.7(\dtau-0.5)\big)} \, .\label{eq:psi_dtau}
	\end{align}
	The function 
	\begin{equation}
		\alpha(\tau, \dtau) = \beta\Psi(\dtau) + (1 - \beta)\Phi(\tau) 
	\end{equation}
	combines the two models into one joint probability and can be interpreted as likelihood that a pedestrian will cross the road given the instantaneous vehicle behavior by means of $\tau$ and $\dtau$. The parameter $\beta$ works as a modulator of the time gap and the vehicle behavior. This modulation allows describing different phases of the interaction through, for example, a variable influence of $\tau$ and $\dtau$ depending on time or distance. In contrast to pure time gap based prediction models, this model sustains a substantial crossing probability at rather small time gaps if the respective vehicle behavior indicates clear yielding signs. Vice versa, the crossing probability decreases even at large gaps if the vehicle shows clear signs of not yielding. The model parameters have been approximated according to the results of related studies \cite{tian.2023, pekkanen.2022, theisen.2024, petzoldt.2014}. 
	Similarly, 
	\begin{equation}
		\alphamin(\tau, \dtau) = 1-\alpha(\tau, \dtau)
	\end{equation}
	can be interpreted as likelihood that a pedestrian will \textit{not} cross the road given the instantaneous vehicle behavior. 
	
	According to \cite{theisen.2024, kalantari.2023, pekkanen.2022} evidence accumulation models offer a suitable way to account for the decision making process of pedestrians that repeatedly assess the situation and accumulate evidence for yielding behavior over time. Thus, we assume a decision making frequency $\fdm=\frac{1}{\Tdm}$ with which the pedestrian evaluates the vehicle behavior and makes a crossing decision until the pedestrian eventually crosses the road. This leads to the discretized instantaneous likelihood $\alphamind(k)$ of the pedestrian not crossing before the vehicle which has been sampled at the decision making time steps $k\cdot\Tdm$ with $k \in [0, ..., N]$ where $N=\frac{\Tmax}{\Tdm}$ denotes the maximum prediction horizon. Given the planned vehicle trajectory $\xi:[0, \Tmax]\rightarrow\mathbb{R}^n$ the cumulative probability that a pedestrian will have crossed at a certain time $T_m$ in the future evolution of the interaction can be estimated as 
	\begin{align}
		P_{t=T_m}\big(\textrm{cross}\mid \xi, \Tdm\big) &= 1-P_{t=T_m}\big(\textrm{stand}\mid \xi, \Tdm\big) \nonumber \\ 
		&= 1- \exp\Big(\sum_{k=0}^{T_m/\Tdm} \log\big(\alphamind(k)\big) \Big)\,. \label{eq:P_cross}
	\end{align}
	The assumption that the crossing decision can be sampled at discrete time steps reinforces the effect that clear yielding behavior leads to an increasing crossing probability. The sampling requires the yielding cues to last for a certain time such that the pedestrian can perceive the communication. If a vehicle only slightly changes behavior for a very short period of time, chances are high that the pedestrian misses the communication and cannot accumulate enough evidence.
	
	In the proposed model, evidence is expressed as probability assuming values between 0 and 1. Hence, within a single planning step, the accumulated evidence can not decrease. This can be reasoned by the fact that the cumulative crossing probability predicted over the interaction horizon can not be less than the initial crossing probability. Within the scope of closed-loop simulations, the initial probability will be reset according to the states in each simulation step, such that the crossing probability can decrease between two consecutive simulation steps depending on the evolution of the scenario.
	\subsection{Vehicle Behavior Validation Step}
	After solving the two subsequent \acrshort{ocps} for the $\tei$-$\sei$-grid, a suitable trajectory needs to be selected to plan the vehicle behavior. According to \cite{mertens.2020}, cooperative behavior can be described as any behavior that reduces the cost of the other agent. However, ego behavior that is intended to be cooperative requires an appropriate action of the interaction partner in order to achieve the intended cooperative resolution. The authors in \cite{mertens.2020} argue that cooperative interactions require approximating the other agent's reaction if no complete information about one's intention and the planned future behavior is available. At the same time, they argue that any kind of behavior, whether it entails an adjustment of the behavior or not, includes communication to some extent. Hence, the ego behavior needs to be carefully selected regarding the information that is implicitly communicated to the other agent. Following this argument, we perform a validation step estimating the pedestrian's reaction to a certain vehicle behavior to confirm that the planned behavior will cause the expected reaction. We use a joint cost function that evaluates each feasible trajectory with respect to the vehicle's comfort and utility, and which takes the estimated utility of the pedestrian into account. Comfort is approximated by means of the vehicle jerk $j$, similar to \eqref{eq:objective_fun}
	\begin{equation}
		\ccomfv = \int_{t_0}^{\te} \frac{\wj}{2}j^2 \textrm{d}t \label{eq:C_comf_veh} \, .
	\end{equation}
	Equation \eqref{eq:C_util_veh} describes the vehicle utility benefit, where $\wTBv$ denotes a weighting factor for the travel benefit and $\te$ denotes the time that the vehicle needs to pass the pedestrian
	\begin{equation}
		\cutilv = \frac{d(0)}{\te v(0)} \wTBv \int_{t_0}^{\te}v\textrm{d}t \label{eq:C_util_veh} \, .
	\end{equation}
	This formulation can be interpreted as a weighted benefit of the covered distance normalized with respect to the time the vehicle would need to pass if it drove with \acrfull{cv}. Since we can only approximate the pedestrian behavior, we propose to consider the pedestrian's costs probabilistically to account for the uncertainty of the actual behavior. The probabilistic formulation of the pedestrian utility is given by 
	\begin{multline}
		\cutilp = \frac{d(0)}{\te v(0)} \wTBp \int_{t_0}^{\te}\vped P\big(\textrm{cross}\mid \xi \big) \textrm{d}t + \\
		\underbrace{\wWTp \int_{t_0}^{\te}P\big(\textrm{stand}\mid \xi \big) \textrm{d}t}_{=\Twt} \label{eq:C_util_ped} \, .	
	\end{multline}
	
	The first part of \eqref{eq:C_util_ped} describes the benefit that the pedestrian gets by walking with an assumed constant speed $\vped$, which is multiplied with a weight $\wTBp$. Moreover, the term in the integral is weighted with the cumulative probability of the pedestrian crossing such that the utility is proportional to the likelihood that the pedestrian has crossed. Additionally, pedestrian utility contains a term that considers the estimated waiting time $\Twt$ of the pedestrian, which is given by the integral of the cumulative standing probability over the whole time horizon of the interaction. This term is weighted with a factor $\wWTp$. If the pedestrian is unlikely to cross before the vehicle given the planned behavior, the cumulative standing probability is high throughout the whole interaction, leading to a large estimated waiting time (with an upper bound given by the time the vehicle needs to pass the pedestrian). Similarly, if the pedestrian is likely to cross at an early stage, the cumulative standing probability is low, resulting in a low estimated waiting time. Eventually, a vehicle behavior is selected such that it leads to the lowest joint costs
	\begin{equation}
		\cjoint = \ccomfv + \cutilv + \cutilp \, . \label{eq:C_joint}
	\end{equation}
	\section{Example Interactions}
\begin{figure*}[!th]
		\centering
		\begin{minipage}{0.98\textwidth}
			\centering
			\includegraphics[width=0.99\textwidth]{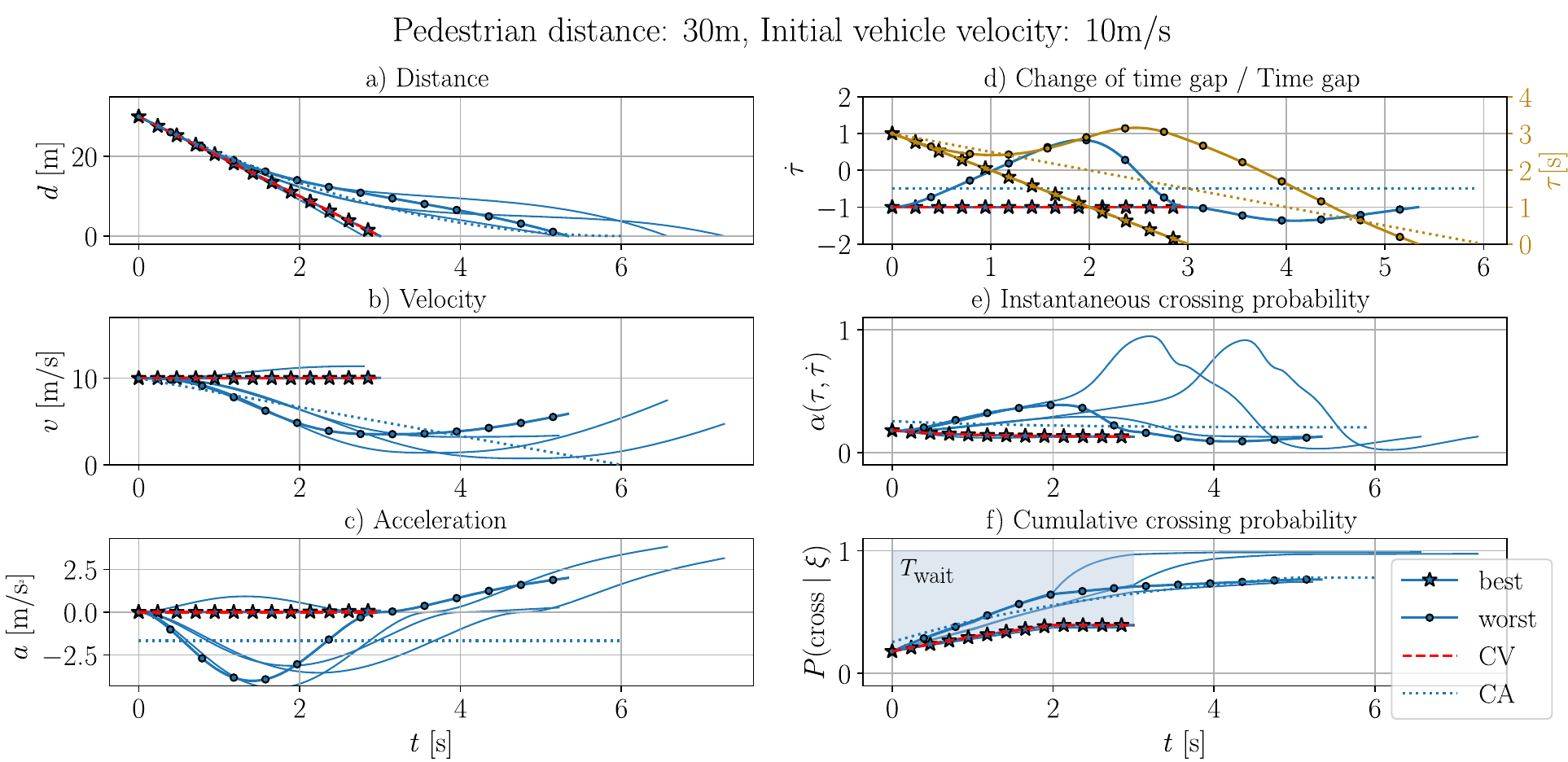}
		\end{minipage}
		\caption{Results of behavior planning and validation at starting distance $30\textrm{m}$ with initial vehicle velocity $10\frac{\textrm{m}}{\textrm{s}}$. For these starting conditions, the vehicle decides that driving on with constant velocity is the best option. The vehicle predicts that this behavior is interpreted as not yielding by the pedestrian.}
		\label{fig:30m}
		\vspace*{-3mm}
	\end{figure*}

	In this section, we analyze the performance of the proposed planning approach in several exemplary situations at different initial vehicle positions and discuss the influence of the validation step on the decision making of the vehicle. We assume a set of values for the parameters of the optimization and evaluation of the vehicle behavior as shown in \tabref{tab:params}. Note that the concept and implications presented in this work are not affected by the exact numbers and remain valid also for different sets of parameters. For the experiments, we use a constant $\beta$ which is fitted according to the best model of \cite{pekkanen.2022}.
	\begin{table}[t]
		\vspace*{-3mm}
		\caption{Optimization and pedestrian behavior estimation parameters}
		\begin{center}
			\begin{tabular}{ |l| l |l |}
				\hline
				\textbf{Parameter}  & \textbf{Value} & \textbf{Description}  \\
				\hline
				$s_0$  & $0\mathrm{m}$ & Initial vehicle position \\
				\hline
				$v_0$  & $10\mathrm{m}/\mathrm{s}$ & Initial vehicle velocity \\
				\hline
				$a_0$  & $0\mathrm{m}/\mathrm{s^2}$ & Initial vehicle acceleration \\
				\hline
				$j_0$  & $0\mathrm{m}/\mathrm{s^3}$ & Initial vehicle jerk \\
				\hline
				\sped  & $30, 40, 90\mathrm{m}$ & Pedestrian position \\
				\hline
				\vped  & $1.5\mathrm{m}/\mathrm{s}$ & Assumed pedestrian walking speed \\
				\hline
				\Tdm  & $1\textrm{s}$ & Assumed pedestrian sampling rate \\
				\hline
				\temax & $10\textrm{s}$ & Largest end time for sampling grid \\
				\hline
				$\Delta\sei$ & $1\textrm{m}$ & Step size of end position sampling \\
				\hline
				$\Delta\tei$ & $0.2\textrm{s}$ & Step size of end time sampling \\
				\hline
				\wj  & $2.25\mathrm{e}{-04}$ & Weight for jerk penalization \\
				\hline
				\wu  & $1.8\mathrm{e}{-04}$ & Weight for control input penalization \\
				\hline
				\wte  & $3\mathrm{e}{-03}$ & Weight for end time penalization \\
				\hline
				\wTBv  & $-3\mathrm{e}{-04}$ & Weight for vehicle traffic benefit \\
				\hline
				\wTBp & $-1.4\mathrm{e}{-02}$ & Weight for pedestrian traffic benefit\\
				\hline
				\wWTp & $5\mathrm{e}{-02}$ & Weight for pedestrian waiting time \\
				\hline
				$\beta$ & $0.3711$ & \makecell{Modulation factor for combination \\ of probabilistic models $\Psi(\dtau)$ and $\Phi(\tau)$} \\
				\hline
			\end{tabular}
			\vspace*{-5mm}
			\label{tab:params}
		\end{center}
	\end{table}
	We consider three exemplary situations with a starting distance between vehicle and pedestrian of $30\mathrm{m}$, $40\mathrm{m}$, and $90\mathrm{m}$, respectively. The initial vehicle velocity is $v_0=10\frac{\textrm{m}}{\textrm{s}}$ in all situations. \figref{fig:30m} depicts the results of the interactions with starting distance $30\mathrm{m}$. Plots \figref{fig:30m}a)--c) show the vehicle kinematics, i.e. distance, velocity, and acceleration. \figref{fig:30m}d) depicts the time gap $\tau$ and change rate of time gap $\dtau$. The curves of the instantaneous crossing probability $\alpha(\tau, \dtau)$ which results from the progression of $\tau$ and $\dtau$ are shown in \figref{fig:30m}e). The cumulative crossing probability $P(\textrm{cross}\mid \xi)$ is illustrated in \figref{fig:30m}f). The shaded area in \figref{fig:30m}f) depicts the estimated waiting time $\Twt$, which is calculated according to \eqref{eq:C_util_ped}, for the best trajectory. Each plot shows a selection of possible behavior candidates resulting from the two-stage optimization\footnote[2]{Note that this selection only comprises a fraction of the total amount of optimized trajectories.}. The best and worst behavior candidates with respect to the overall cost evaluation \eqref{eq:C_joint} are highlighted by star and circle markers, respectively. Moreover, the \acrshort{cv} behavior option, which is a result of the optimization for the sampling grid, and a hypothetical \acrfull{ca} trajectory are depicted by dashed and dotted lines, respectively. The \acrshort{ca} trajectory describes a target braking behavior that brings the vehicle to a full stop at the position of the pedestrian $\sped$ with a constant deceleration. 
	
	\begin{figure*}[!th]
		\centering
		\begin{minipage}{0.98\textwidth}
			\centering
			\includegraphics[width=0.99\textwidth]{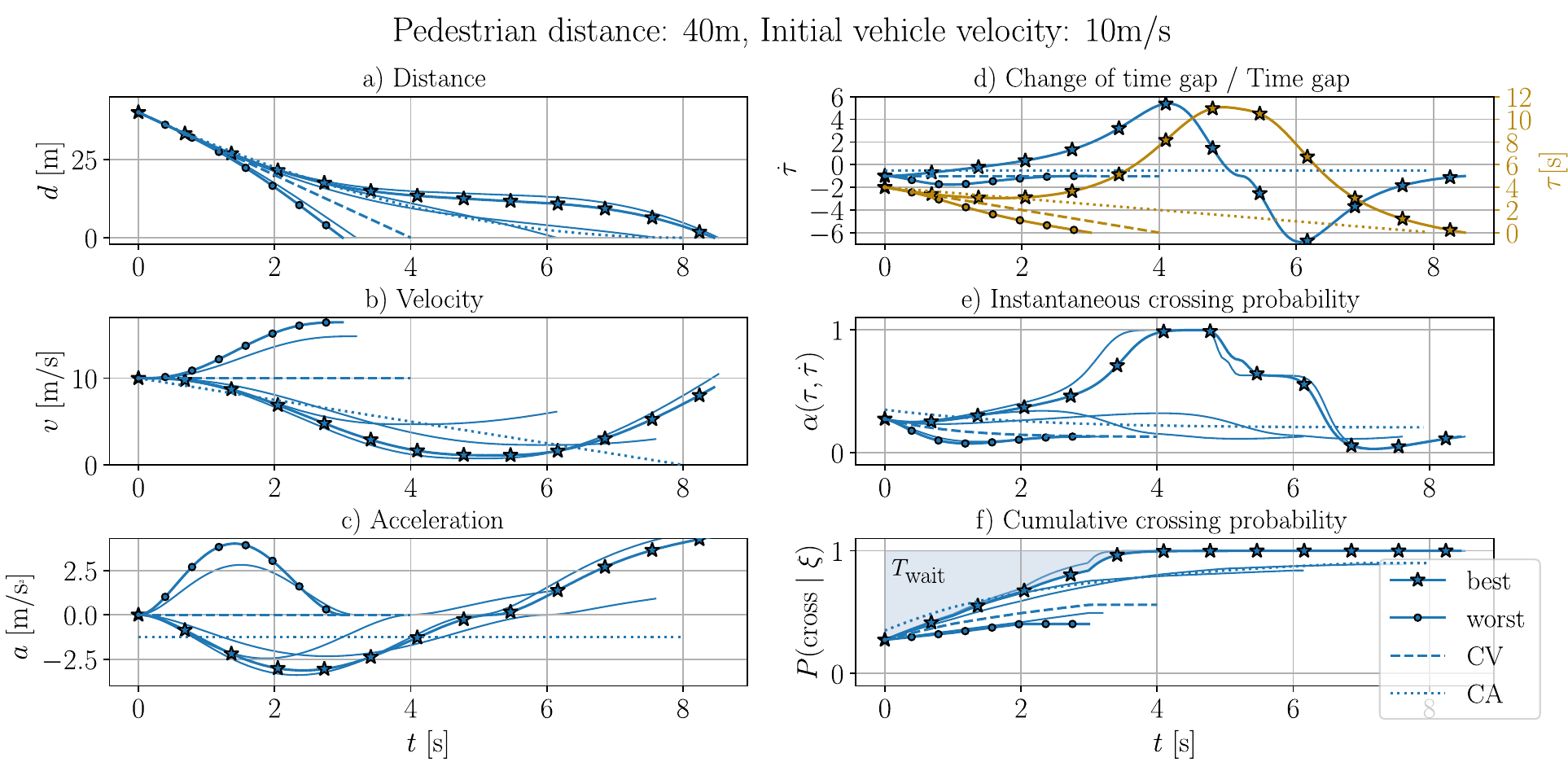}
		\end{minipage}
		\caption{Results of behavior planning and validation at starting distance $40\textrm{m}$ with initial vehicle velocity $10\frac{\textrm{m}}{\textrm{s}}$. The vehicle decides that letting the pedestrian cross is the best option. The vehicle predicts that this behavior requires additional implicit communication to be accepted by the pedestrian. }
		\label{fig:40m}
		\vspace*{-3mm}
	\end{figure*}
	
	\figref{fig:30m}d) shows that the vehicle starts at an initial time gap $\tauinit=3\textrm{s}$ and all trajectories start with $\dtau=-1$, except for \acrshort{ca}. For the \acrshort{ca} candidate, $\dtau$ assumes a constant value of $-\frac{1}{2}$, however, this behavior is only shown for comparison since the trajectory requires a constant non-zero acceleration which does not comply with the initial state $a_0=0$. According to the behavior and gap acceptance model presented in \secref{subsec:pedestrian_model}, the initial crossing probability is low in a situation with a small initial time gap and without any apparent yielding signs. This is illustrated in \figref{fig:30m}e) and \figref{fig:30m}f). It can be seen that the \acrshort{cv} trajectory would be selected as the most cost efficient behavior. The system decides not to yield for the pedestrian in this situation. The system prioritizes vehicle utility benefit and saved discomfort over an increase in crossing probability. This selection can be explained by analyzing the instantaneous and cumulative crossing probabilities. The estimated crossing probability is low over the course of the whole interaction if the vehicle proceeds with \acrshort{cv}. At the same time, the estimated waiting time is low since the system infers that the pedestrian can cross latest when the vehicle has passed. As can be seen in \figref{fig:30m}c), the worst behavior candidate comprises temporary, yet strong, braking. This behavior causes substantial discomfort and utility loss for the vehicle. However, this behavior does not lead to a significant increase of crossing probability. Thus, despite the strong deceleration, the instantaneous yielding cues of the worst option are not clear enough. The uncomfortable and inefficient driving combined with the ambiguous pedestrian reaction leads to the assessment that this behavior is the worst option. The \acrshort{ca} option (dotted lines) would have not been the selected behavior either. Given the assumed sampling rate $\Tdm$ it takes a long time for the pedestrian to accumulate enough evidence that the vehicle is yielding.
	\begin{figure*}[t]
		\centering
		\begin{minipage}{0.98\textwidth}
			\centering
			\includegraphics[width=0.99\textwidth]{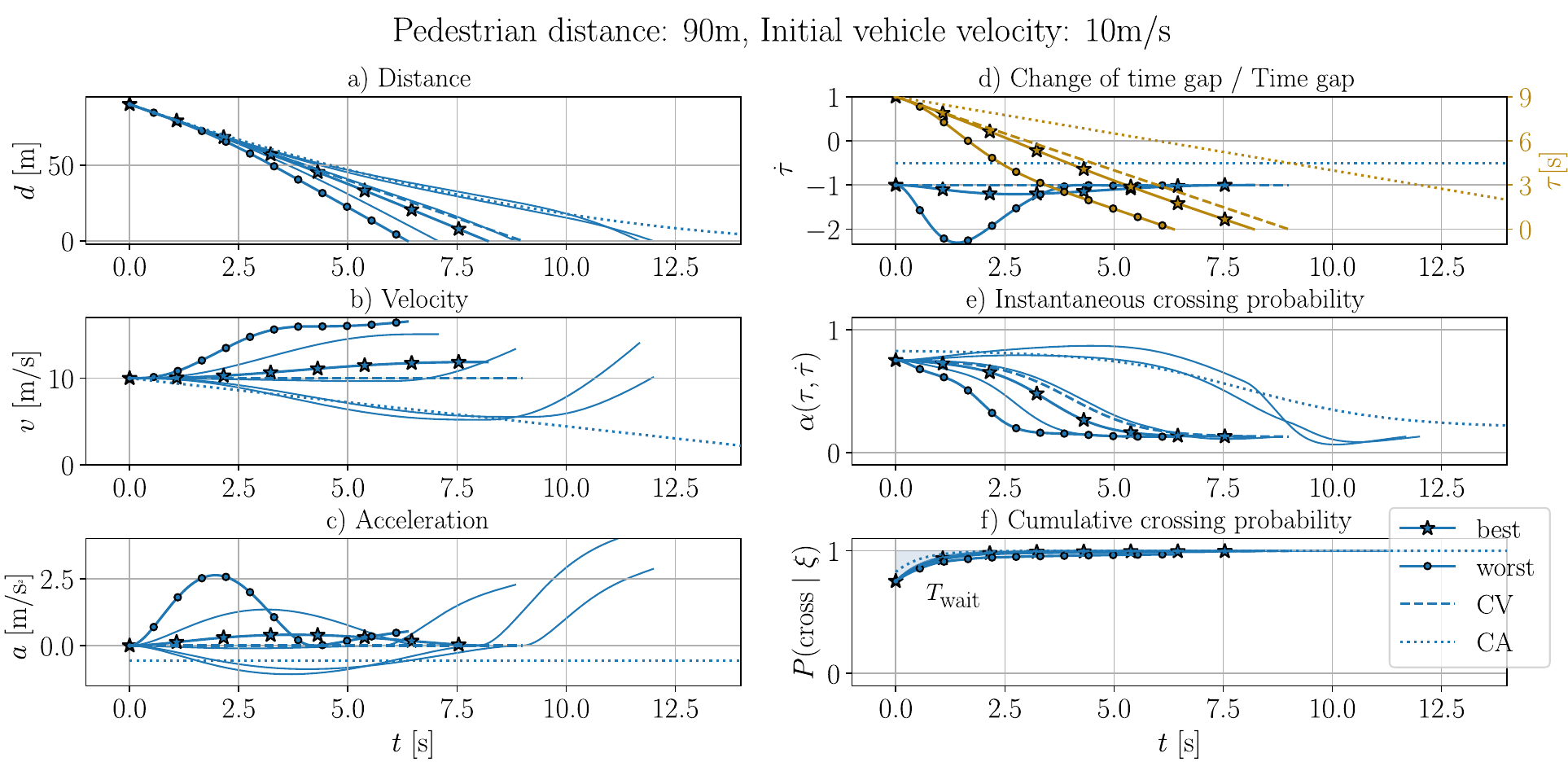}
		\end{minipage}
		\caption{Results of behavior planning and validation at starting distances of 90m with an initial vehicle velocity $v_0=10\frac{\textrm{m}}{\textrm{s}}$. For these starting conditions, no additional yielding cues are required and the best option is to drive on.}
		\label{fig:90m}
		\vspace*{-3mm}
	\end{figure*}
	The previous analysis emphasizes the necessity for validating the ego behavior with respect to the reaction of the interaction partner. Even if the vehicle is yielding comfortably, the cues perceived by the pedestrian might be insufficient to accept the behavior. This can lead to unexpected reactions resulting in inefficient resolution.
	
	\figref{fig:40m} shows the results of the planning and validation for a scenario with the same initial conditions except that the vehicle starts at $40\textrm{m}$ distance. This situation illustrates the shift from deciding to drive through to yielding for the pedestrian. In \figref{fig:40m}d), it can be seen that the selected behavior temporarily reaches values  $\dtau>0$ opening up a large time gap which the system assumes to be sufficient for the pedestrian to cross. Similar trajectories were also created in the 30m scenario, however, due to the parameterization of the joint costs, now, the system decides to accept a certain discomfort and decrease of utility to improve the overall costs by letting the pedestrian cross in this scenario. 
	
	\begin{figure}[t]
		\centering
		\includegraphics[width=0.99\linewidth]{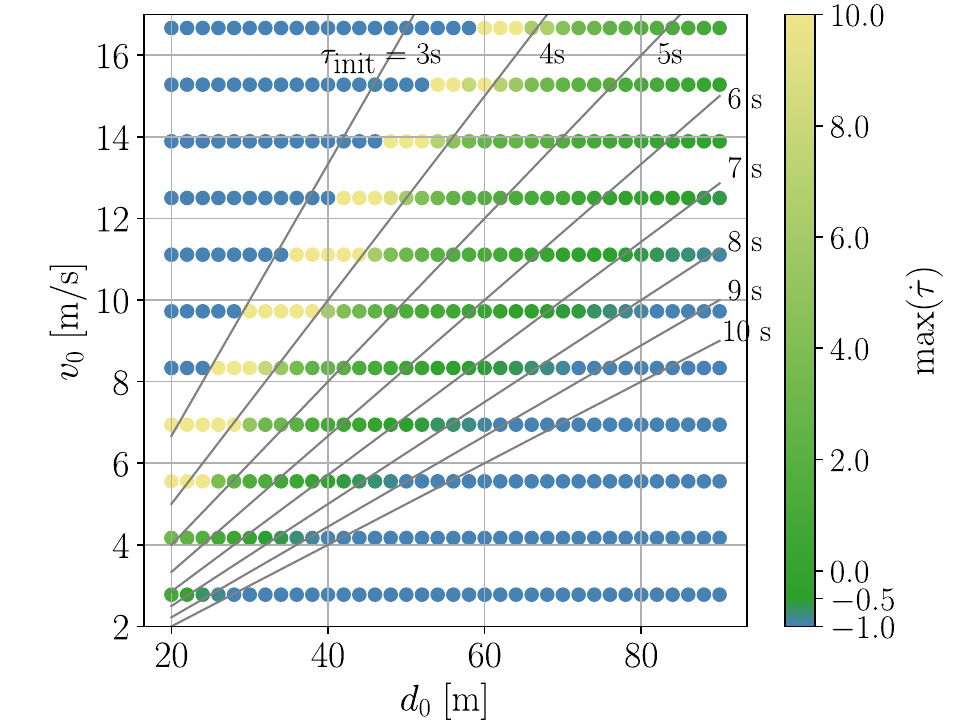}
		\caption{Amount of implicit communication conveyed by the vehicle in a variety of interactions with different initial conditions. Implicit communication is denoted as maximum value of $\dtau$ that is reached by the selected behavior in each situation. The straight lines depict constant time gaps.}
		\label{fig:different_inits_dtau}
		\vspace*{-5mm}
	\end{figure}
	
	\figref{fig:90m} shows the decision making at a starting distance of $90\textrm{m}$. It can be seen that the system assumes that the initial time gap is large enough for the pedestrian to cross. As depicted in \figref{fig:90m}f), the initial crossing probability is approximately $75\%$ and after $1.5\textrm{s}$ the pedestrian could accumulate enough evidence to accept the gap independent of the vehicle behavior. The estimated waiting time is close to zero for all behavior options since the system expects the pedestrian to cross shortly after the start of the interaction. Hence, the selected behavior is mainly influenced by the vehicle costs such that the vehicle slightly accelerates to be as quick as possible without a significant increase of discomfort. 
	
	The previous results underline that the required communication by means of $\dtau$ depends on the starting conditions of a situation. In situations, in which the vehicle has a sufficient yielding margin, letting the pedestrian cross before the vehicle is selected as the most beneficial outcome for the interaction. However, in certain situations, the vehicle's yielding intention might not be apparent to the pedestrian. Such ambiguous situations need particular implicit behavior to communicate the vehicle's intention. \figref{fig:different_inits_dtau} depicts the required amount of implicit yielding communication for a variety of different starting distances and initial velocities. The amount of communication is denoted as the maximum value of $\dtau$ that is reached by the corresponding selected behavior in each set of starting conditions. The blue dots depict behavior with $\max(\dtau)=-1$, which means that the vehicle does not convey any implicit communication cues of yielding for the pedestrian. This is the case in situations with a small initial time $\tau_\textrm{init}\approx 3\textrm{s}$. In these situations, the vehicle decides that it is more beneficial to drive through without yielding. As discussed in the example shown in \figref{fig:30m}, the system predicts a low crossing probability, thus, the vehicle's intention of not stopping is assumed to be evident to the pedestrian. As $\tau_\textrm{init}$ increases, it becomes more beneficial to let the pedestrian cross before the vehicle. However, pronounced implicit communication is necessary to convey the vehicle's intention and induce crossing behavior in ambiguous situations with still rather small time gaps. The additionally required implicit communication declines for larger time gap since the initial crossing probability increases. Note that for very large gaps, the selected behavior is similar to \acrshort{cv} behavior and no additional implicit communication is necessary. If a pedestrian rejects a large gap, leading to incorrectly predicted crossing, the situation evolves and will eventually result in a similar situation at a smaller distance. At this point, the vehicle initiates behavior with stronger implicit communication depending on the new initial states. Thus, the system is able to adapt to this mismatch. 
	
	\section{Limitations and Future Work}
	The prediction models $\alpha$ and $\alphamin$ of the instantaneous crossing and standing probability reflect the influence of the vehicle behavior based on statistical distributions over a broad spectrum of possible reactions. Yet, these models do not represent the specific crossing preferences of an individual pedestrian. Since $\alpha$ and $\alphamin$ reflect the influence of the \textit{instantaneous} vehicle behavior, $\alpha(t+1)$ and $\alphamin(t+1)$ are independent of $\alpha(t)$ and $ \alphamin(t)$, respectively. Thus, at each evaluation time step the model treats the pedestrian's crossing decision as if it is the first time for the pedestrian to make this decision, independent of previous decisions. 
	
	Although the suggested model is based on results from related empirical studies with human participants, the results presented in this work are based on a theoretical model derived from these studies. Hence, the results should be considered as preliminary regarding the effectiveness and the exact parametrization of the model. In future work, we intend to test the planning approach and validate the model in real-world interactions under controlled conditions. Moreover, we plan to extend the model to settings in which the interaction partners are not limited to approaching each other perpendicularly and add behavior acceptance models for other user types such as human drivers. 
	\section{Conclusion}
	In this work, we present a combined prediction and planning approach for vehicle-pedestrian interactions comprising a probabilistic pedestrian behavior acceptance model. The proposed approach incorporates a \textit{validation step} to evaluate whether a vehicle behavior is likely to cause a certain pedestrian reaction. We demonstrate the performance of the approach with the help of exemplary vehicle-pedestrian interactions with different starting conditions. The proposed approach can be deployed as an extension to existing behavior planners to estimate pedestrian reactions and validate ego behavior. This approach will allow future automated vehicles to explicitly control their implicit communication and lead to less uncertainty in interactive traffic situations.


	
	\newacronym{av}{AV}{automated vehicle}
	\newacronym{avs}{AVs}{automated vehicles}
	\newacronym{cv}{CV}{constant velocity}
	\newacronym{ca}{CA}{constant acceleration}
	\newacronym{hmi}{HMI}{human-machine-interface}
	\newacronym{hmis}{HMIs}{Human-machine-interfaces}
	\newacronym{ehmi}{eHMI}{external human-machine-interface}
	\newacronym{ehmis}{eHMIs}{external human-machine-interfaces}
	\newacronym{ehmis_med}{eHMIs}{external HMIs}
	\newacronym{ihmi}{iHMI}{internal human-machine-interface}
	\newacronym{ihmis}{iHMIs}{internal human-machine-interfaces}
	\newacronym{ocp}{OCP}{optimal control problem}
	\newacronym{OCP}{OCP}{Optimal Control Problem}
	\newacronym{ocps}{OCPs}{optimal control problems}
	\newacronym{vru}{VRU}{vulnerable road user}
	\newacronym{vrus}{VRUs}{vulnerable road users}
	\newacronym{sts}{STS}{shared traffic space}
	\newacronym{tta}{TTA}{time-to-arrival}
	
	\bibliographystyle{IEEEtran}
	\bibliography{./mybib}
	
\end{document}